\documentclass[%
 reprint,
superscriptaddress,
 amsmath,amssymb,
 aps,
]{revtex4-2}

\usepackage{graphicx}
\usepackage{dcolumn}
\usepackage{bm}
\usepackage{braket}
\usepackage{mathrsfs}  
\DeclareMathOperator{\Tr}{Tr}
\usepackage{multirow}
\usepackage{relsize}
\usepackage[colorlinks=true, linkcolor=blue, citecolor=blue, urlcolor=black
]{hyperref}%

\begin{document}

\preprint{APS/123-QED}

\title{Clustering and enhanced classification using a hybrid quantum autoencoder}

\affiliation{School of Physics, University of Melbourne, VIC, Parkville, 3010, Australia.}
\affiliation{School of Mathematics and Statistics, University of Melbourne, VIC, Parkville, 3010, Australia.}


\author{Maiyuren Srikumar}
\email{srikumarm@unimelb.edu.au}
\affiliation{School of Physics, University of Melbourne, VIC, Parkville, 3010, Australia.}

\author{Charles D. Hill}
\email{cdhill@unimelb.edu.au}
\affiliation{School of Physics, University of Melbourne, VIC, Parkville, 3010, Australia.}
\affiliation{School of Mathematics and Statistics, University of Melbourne, VIC, Parkville, 3010, Australia.}
\author{Lloyd C.L. Hollenberg}
\email{lloydch@unimelb.edu.au}
\affiliation{School of Physics, University of Melbourne, VIC, Parkville, 3010, Australia.}

\date{\today}

\begin{abstract}

Quantum machine learning (QML) is a rapidly growing area of research at the intersection of classical machine learning and quantum information theory. 
One area of considerable interest is the use of QML to learn information contained within quantum states themselves. In this work, we propose a novel approach in which the extraction of information from quantum states is undertaken in a classical representational-space, obtained through the training of a \textit{hybrid quantum autoencoder} (HQA). Hence, given a set of pure states, this variational QML algorithm learns to identify -- and classically represent -- their essential distinguishing characteristics, subsequently giving rise to a new paradigm for clustering and semi-supervised classification. The analysis and employment of the HQA model are presented in the context of amplitude encoded states -- which in principle can be extended to arbitrary states for the analysis of structure in non-trivial quantum data sets.

\footnotetext{
$^\dag$srikumarm@student.unimelb.edu.au\\
$^\ddagger$cdhill@unimelb.edu.au\\
$^\S$lloydch@unimelb.edu.au}

\end{abstract}

\maketitle


\section{\label{sec:level1}Introduction} 

In recent years, the amalgamation of Quantum Mechanics and Machine Learning (ML) has instigated extensive research into the field of Quantum Machine Learning (QML) \cite{10.1038/nature23474}. 
With fault-tolerant quantum computation far from realisable in the near future, one of the areas in which researchers are looking for quantum advantage are variational algorithms. These variational approaches have demonstrated robustness in the regime of noisy intermediate-scale quantum (NISQ) devices, thus being a contender to first demonstrate quantum advantage \cite{10.22331/q-2018-08-06-79}. With applications in QML variational methods most commonly employ a parameterised quantum circuit (PQC) \cite{10.1038/ncomms5213, 10.1088/1367-2630/18/2/023023}, where parameters are classically optimised in a feedback loop routine between optimiser and PQC. 

\begin{figure}
    \centering
    \includegraphics[width=246pt]{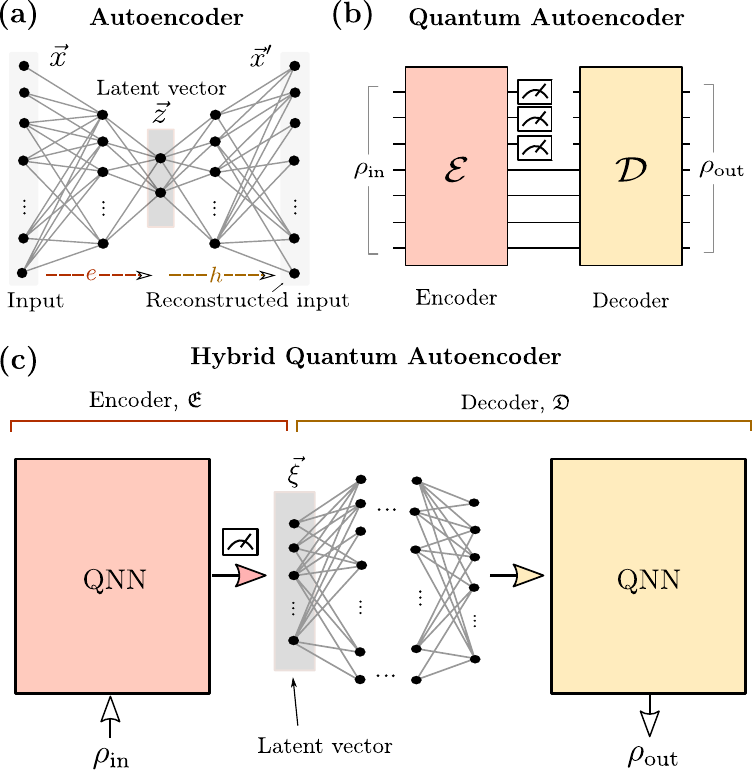}
    \vspace{-0.4cm}
    \caption{(a) An autoencoder, where black dots represent \textit{neurons}, and lines represent their weighted inputs/outputs (Appendix \ref{sec:ann}). (b) Structure of a quantum autoencoder as initially proposed in \cite{10.1088/2058-9565/aa8072}. $\mathcal{E}$ and $\mathcal{D}=\mathcal{E}^{\dagger}$ are unitary operators. Here the latent vector is an inaccessible compressed quantum state. (c) Structure of the \textit{hybrid quantum autoencoder} (HQA). The encoder is composed of a quantum neural network (QNN) that is measured to return a \textit{classical} latent vector, $\vec{\xi}$. The decoder employs an ANN and a QNN to return quantum state. The output of all autoencoders, (a), (b) \& (c) are trained to match any input from a particular data set.  }
    \label{fig:general_hqa_fig}
\end{figure}

Generally, QML methods can be categorised into two distinct groups: (i) models that obtain advantage through the learning of \textit{classical} data -- once embedded into a quantum system -- or (ii) the learning of purely \textit{quantum} data sets. This paper focuses on the latter task, and proposes an approach in which the learning of quantum states can be undertaken in a classical representational space. This allows for novel approaches to cluster and classify quantum states based on their classical representations. 
Such representations will be formed from the employment of -- what we have termed -- a \textit{hybrid quantum autoencoder} (HQA), illustrated in Figure \ref{fig:general_hqa_fig}c.

Classically, an \textit{autoencoder} is a specific artificial neural network (ANN) architecture that is trained to return its input as its output, whilst undergoing a crucial funnelling of its degrees of freedom \cite{10.5555/3086952} (shown in Figure \ref{fig:general_hqa_fig}a). This funnelling process generates compressed representations of data points that belong to a particular group of data.

The autoencoder is composed of two maps: an \textit{encoder}, $e$, and \textit{decoder}, $h$, that are both approximated using ANNs (a discussion on ANNs relevant to this work is presented in Appendix \ref{sec:ann}). The encoder receives data $\vec{x}$ and outputs a lower dimensional \textit{latent vector}, $\vec{z}$, such that, $\vec{z}=e(\vec{x})$. The decoder performs the inverse, $\vec{x}' = h(\vec{z})$. Subsequently, the autoencoder (both $h$ and $e$) is trained to approximate $\vec{x} \approx h(e(\vec{x}))$ for any $\vec{x}$ in  the data set. This would be trivial for if $\vec{x}$ and $\vec{z}$ had equal dimensionality; however, in the case where  $\text{dim}(\vec{x})>\text{dim}(\vec{z})$, the autoencoder is forced to encode the most important aspects of the input, $\vec{x}$, into the latent space. The \textit{latent vector}, $\vec{z} \in \mathcal{Z}$, is in essence a representation of $\vec{x} \in \mathcal{X}$ in the lower dimensional representational space of $\mathcal{Z}$. 

The main advantage of an autoencoder is that it is able to learn complex compression strategies through an unsupervised learning process. Such a process requires a human to have minimal prior information regarding the data set. Hence, it is often used in the context of denoising and compressing data that lack obvious methods of dimensionality reduction. Quantum autoencoders (Figure \ref{fig:general_hqa_fig}b), discussed further in Section (\ref{sec:qae}), are direct analogues and hence provide non-trivial compression maps for quantum states to a subset of its Hilbert space.

In essence, both ML and QML algorithms exploit the tendency that data aggregates in a low dimensional sub-manifold over the vast space of possible data points. 
We describe the data as lying on a sub-manifold to emphasise the fact that infinitesimal tangential translations, result in remaining on the sub-manifold. It should be noted that, although in mathematics a \textit{manifold} has a more formal definition, in ML it is used to describe a set of points that can be well approximated by considering only a small number of degrees of freedom, embedded in a higher-dimensional space \cite{10.5555/3086952}. This \textit{manifold hypothesis} is essential so that data points have a neighbourhood of highly similar examples that can be accessed by applying small transformations to traverse the manifold. Hence, the main objective of a quantum autoencoder is to learn the sub-manifold that describes a particular set of quantum states. The HQA in particular, represents the sub-manifold in an accessible classical vector space.

Quantum states can be represented as positive semi-definite operators on a complex Hilbert space, $\mathbb{C}^{2^n}$, known as density matrices. Theoretically, one can imagine putting the elements of the density matrix through a classical ML algorithm to find similarities between quantum states or even classify states. However, the information stored in quantum states is notoriously inaccessible without exponential resources to characterise each (through quantum state tomography \cite{10.1103/physrevlett.105.150401, 10.1088/1367-2630/14/9/095022}). The HQA, in Figure \ref{fig:general_hqa_fig}b, gets around this by using an \textit{encoder} that is trained to output classical information about important aspects of a quantum state. This quantum to classical transformation is extreme in its dimensionality reduction, as for a pure input state we have $\mathbb{C}^{2^n} \xrightarrow[]{} \mathbb{R}^{v}$ where $v\sim \mathcal{O}(n)$ is the dimension of the classical space by measuring the $n$ number of qubits. Hence one can see that if the state is able to be reconstructed from the classical space, then the classical space can only describe a relatively small set of quantum states. Nonetheless, the classical real space \textit{represents} a manifold in $\mathbb{C}^{2^n}$ that can be learned with the construction of the HQA. It will be seen that it is this perspective that distinguishes the HQA from the QAEs explored in literature thus far.

This paper is structured such that we first provide background into quantum neural networks and quantum autoencoders in Section (\ref{sec:background});  before then constructing the HQA in Section (\ref{sec:HQA}). This is then followed by applications of clustering and classification in Section (\ref{sec:application}), including results from numerical simulations.

\section{\label{sec:background}Background}

\subsection{Quantum Neural Networks}

Quantum neural networks (QNNs) are the extension of ANNs to QML. The precise form of the QNN, however, is quite non-trivial as it would need to take advantage of unique quantum mechanical properties while also retaining the non-linear functional features of classical ANNs \cite{10.1007/s11128-014-0809-8}. Hence, there are various proposals for QNN designs that claim similar non-linear dissipative dynamics of ANNs, but are yet to present clear quantum advantage \cite{10.1038/s41534-017-0032-4, 10.1103/physreva.98.042308, 10.1103/physreva.100.012334, Cao_Quantum_Neuron}.
In this paper, the implementation of the HQA will use the simplest design of a QNN: a PQC coupled with a specified observable. The choice of QNN, however, is arbitrary to the overall approach of the HQA (as shown in Figure \ref{fig:general_hqa_fig}c). Hence a fair comparison of QNN complexity and expressibility -- for the construction of the HQA specifically -- is left for future work.

PQCs form the basis of hybrid quantum-classical algorithms that optimise a quantum circuit with respect to a problem-dependent cost function \cite{10.1088/2058-9565/aab822}. The optimisation is performed classically to determine a better estimate of parameters which define a variational circuit. The optimisation in this work is performed using the \textit{parameter shift rule} \cite{10.1103/physreva.99.032331}, elaborated in Appendix \ref{sec:param_shift}. 

Since any quantum circuit can be defined as a gate sequence $\mathcal{U}(\theta)$, the $m$ parameters of the circuit are the set of $\theta_i$ which parameterise the unitaries. We define the measurement of a PQC as a function $f : \mathbb{R}^m \xrightarrow{} \mathbb{R}$, mapping the gate parameters to an expectation value, 
\begin{equation}\label{eq:variational_circuit}
    f_{\rho_0}(\theta) := \Tr \big[\mathcal{U}(\theta)\rho_0 \mathcal{U}^{\dagger}(\theta)\hat{B}\big] := \Tr \big[\rho(\theta) \hat{B}\big],
\end{equation}
\noindent where $\hat{B}$ is a predetermined observable (most commonly a Pauli-Z) and $\rho_0$ is an initial arbitrary state of the circuit.  For consistency, the state before measurement will have notation, $\rho(\theta)= \mathcal{U}(\theta)\rho_0 \mathcal{U}^{\dagger}(\theta)$  as evident from the second line in equation \eqref{eq:variational_circuit}.
It should be noted that this functional form of the variational circuits hides the fact that on real devices, repeated measurements of the circuit are required to obtain this expectation value. As a result, there is a natural statistical uncertainty to the \textit{estimated} $f$, determined by the number of samples taken of the circuit. Caution is thus required when constructing algorithms that require arbitrarily large numerical precision of $f$. These algorithms may have promising results in state-vector simulations, but become computationally infeasible to measure on real quantum devices, requiring an exponentially large number of samples.

The optimisation of PQCs remaim an area of research as it is not clear that employing classical optimisers will result in optimal solutions for quantum cost function landscapes. This has given rise to works suggesting optimisers that are aware of the underlying quantum structure of quantum states \cite{10.22331/q-2020-05-11-263, 10.22331/q-2020-08-31-314, 10.22331/q-2020-05-25-269}. Furthermore, there exist limitations of \textit{barren plateaus} in deep PQC-based algorithms, as initially realised in \cite{10.1038/s41467-018-07090-4}. Here, the exponential suppression of gradient with increasing depth of circuit, has also been shown to be linked to the expressibility of PQCs \cite{holmes2021}. In addition to barren plateaus, PQCs are seen to exhibit \textit{narrow gorges} \cite{10.1038/s41467-021-21728-w} which is the occurrence of the existence of cost landscape minima in narrow wells that get steeper with increasing depth. 
The effects are not only seen with an increase in qubits, but arise due to entanglement \cite{marrero2021entanglement}, certain cost functions \cite{10.1038/s41467-021-21728-w, arrasmith2021equivalence},  and noise \cite{wang2021noiseinduced}.
These phenomena clearly have implications on QAEs that employ PQCs, hence it is important that an implementation of the HQA is \textit{efficient} in its circuit depth -- i.e. $\mathcal{O}(n)$.  This is largely dependent on the chosen quantum data set $\{\rho_{\text{in}}\}$, and the expressibility of the particular QNN component shown in Figure \ref{fig:general_hqa_fig}c.

\subsection{Quantum autoencoders}\label{sec:qae}

Many recent works involving quantum autoencoders (QAEs) build on the structure first proposed by Romero et al. \cite{10.1088/2058-9565/aa8072}, using shallow PQCs to compress quantum states. In \cite{10.1103/physrevlett.124.130502}, QNNs are of the form in \cite{10.1038/s41467-020-14454-2} to construct a QAE that successfully denoises Greenberger-Horne-Zeilinger states which are subject to spin-flip and random unitary noise errors. In \cite{10.1088/2058-9565/aae22b}, a QAE is constructed using approximate quantum adders that are obtained with classical genetic algorithms as opposed to more commonly used gradient methods for parameter optimisation.

All such applications work on the process of funnelling quantum states into a lower dimensional Hilbert space. This naturally returns compressed representations that disregard both stochastic noise fluctuations and irrelevant degrees of freedom. It is important to note that the set of states are assumed to have support on a subset of its Hilbert space, $\mathcal{S} \subset \mathcal{H}$. The existence of such support is not guaranteed, but is instead common in many sets of quantum states, due to symmetries inherent to physical processes. For example, in \cite{10.1088/2058-9565/aa8072} a QAE is classically simulated to show the compression of ground states of the Hubbard model and molecular Hamiltonians. 

QAEs have been experimentally realised for the compression of qutrits with photons in \cite{10.1103/physrevlett.122.060501} and the compression of two-qubit states into two single-qubit states in \cite{10.1103/physreva.102.032412}. Furthermore experimental realisation of the QAE via quantum adders has been shown in \cite{10.1002/qute.201800065}. These QAEs are promising, but are nonetheless distinct to the HQA proposed in this paper. Specifically, the HQA aims to generate a \textit{classical} representation of the $\mathcal{S}$ sub-manifold for classical analysis. Hence, not only can the data be compressed, but the compressed representation is an accessible classical vector that can be analysed. 

There are clear limitations on an autoencoder's ability to compress data. On the compression rate of QAEs, there exist not only a fundamental limit due to the degrees of freedom in a dataset, but a quantum limitation related to the von Neumann entropy of the density operator representing the ensemble of training states \cite{10.1088/2058-9565/aa8072}. In \cite{ma2020compression}, it is further elaborated that the compressibility is related to the eigenvalues of the weighted ensemble density matrix. Crucially, this theoretical limit is intrinsic to all possible compression strategies and QAEs. 

\section{The Hybrid Quantum Autoencoder}\label{sec:HQA}

\subsection{Design}\label{sec:hqa_design}

\begin{figure*}
\includegraphics[width=16cm]{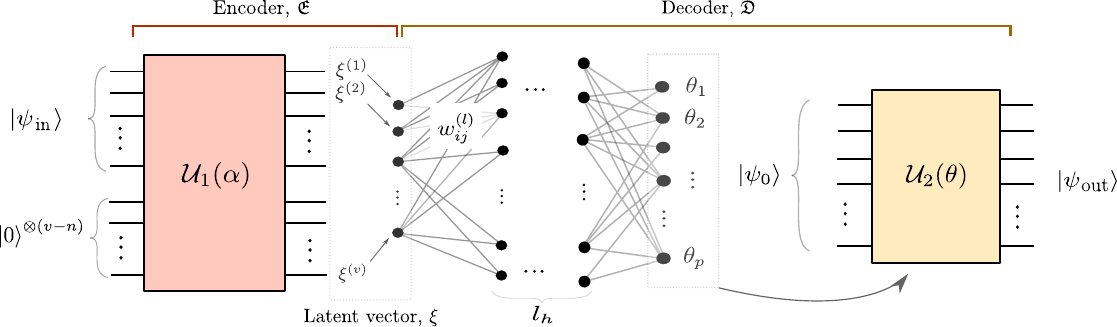}
\caption{Illustration of the PQC-based hybrid quantum autoencoder (HQA) constructed through the combination of an encoder, $\mathfrak{E}$, and decoder, $\mathfrak{D}$. With input $\ket{\psi_{\text{in}}} \in \mathcal{H}_2^{\otimes n}$, there are $(v-n)$ ancilla qubits. The latent space in the diagram is defined as the vector of dimension $v$, formed from the expectation values of all qubits in register, $\xi=(\xi^{(1)}, ..., \xi^{(v)})$, where $\xi^{(i)}=\langle Z_i \rangle$ and $\text{Z}_i$ is the Z-pauli operator acting on the $i^{\text{th}}$ qubit. The PQC architecture of both $\mathcal{U}_{1}$ and $\mathcal{U}_{2}$ are shown in Appendix \ref{sec:pqc_arch}. }
    \label{fig:hqa_figure}
\end{figure*}

This paper proposes a novel variation of the QAE that we have termed a \textit{Hybrid Quantum Autoencoder} (HQA). The hybrid nature of this model arises from the incorporation of both ML, in the form of classical ANNs, as well as QML, through the use of PQC-based QNNs. Figure \ref{fig:hqa_figure} illustrates the overall design of the model which is a combination of (i) an \textit{encoder} that takes a quantum state from Hilbert space $\mathcal{H}^{\otimes n}_2$ to a subset of the real vector space $\mathcal{V}$ of dimension $v=\mathrm{dim}(\mathcal{V})$, and (ii) a \textit{decoder} that performs the inverse of such an operation. In general, though quantum states are positive semidefinite (trace$=1$) operators $\rho$ on $\mathcal{H}^{\otimes n}_2$, the HQA is equipped to identify only pure states. Hence, we associate the vector $\ket{\psi_a}$ to the pure state $\rho_a := \ket{\psi_a}\bra{\psi_a}$.
Mathematically, the encoder and decoder have the form of a map,
 \begin{align}
     \mathfrak{E}: \ & \mathcal{H}^{\otimes n}_2 \xrightarrow{} \mathcal{V} \text{, \ \ \ \ \ where \ \ \ } \mathfrak{E}(\ket{\psi_{\text{in}}})=\xi , \\ 
     \mathfrak{D}: \ & \mathcal{V} \xrightarrow{} \mathcal{H}^{\otimes n}_2 \text{, \ \ \ \ \ where \ \ \ } \mathfrak{D}(\xi)=\ket{\psi_{\text{out}}} ,
 \end{align}
\noindent where $\mathcal{V}=[-1,1]^{v}$ is termed the \textit{latent space} and $\xi \in \mathcal{V}$ is referred to as the \textit{latent vector} -- analogous to the terminology used when dealing with classical autoencoders. This vector is in essence a classical representation of the quantum state -- a \textit{perfect} representation if $\mathfrak{D} \circ \mathfrak{E} (\ket{\psi_{\text{in}}}) = \ket{\psi_{\text{in}}}$ is achieved. This indicates that the information of state $\ket{\psi_{\text{in}}}$ was preserved in the latent vector to then be recreated without any loss of information. In \textit{information theory} this is lossless encoding of information into a compressed latent space. 

Though the functional form of the encoder and decoder are defined, the models themselves have not been specified. As seen in Figure \ref{fig:hqa_figure}, the encoder $\mathfrak{E}$ is a PQC parameterised by a vector, $\alpha$. The PQC receives some state $\ket{\psi_{\text{in}}}$, and applies unitary $\mathcal{U}_1(\alpha)$ on the combined system of the input state with $(v-n)$ ancilla qubits. From this circuit, the Z expectation value of every qubit is measured, to then form the latent vector $\xi$, 
\begin{equation}\label{eq:encoder}
    \mathfrak{E}^{(i)}(\ket{\psi_{\text{in}}}) = \xi^{(i)} = \bra{\widetilde{\psi_{\text{in}}}} \text{Z}_i\mathcal{U}_1(\alpha) \ket{\widetilde{\psi_{\text{in}}}} , 
\end{equation}
\noindent where $\ket{\widetilde{\psi_{\text{in}}}} = \ket{\psi_{\text{in}}} \otimes \ket{0}^{\otimes (v-n)}$, $\text{Z}_i$ is the Z-Pauli operator acting on the $i^{\text{th}}$ qubit and $\alpha$ parameterises the PQC that will be optimised. The set of $\xi = (\xi^{(1)}, ..., \xi^{(v)})$ forms the latent vector which is identified as the classical representation of an input quantum state. Following the encoder, the decoder is extremely similar to the one defined in equation \eqref{eq:encoder}. However, this time it is a mapping that returns a quantum state when given a latent vector, 
\begin{equation}\label{eq:decoder}
    \mathfrak{D}_{A}(\xi) = \mathcal{U}_2\big(f(\xi) \big)\ket{0}\ket{\psi_0} , 
\end{equation}
\noindent where $f: \mathcal{V} \xrightarrow{} \mathbb{R}^{p}$ is the functional from of the ANN that is parameterised by $w_{ij}^{(l)}$ (discussed in Appendix \ref{sec:ann}). Training changes only the weights of the ANN and not the parameters of the PQC. The PQC parameters are designed to be the output of the ANN. $\ket{\psi_0}$ in Eq. \eqref{eq:decoder} is an appropriate $n$-qubit ansatz for the type states involved. In this paper, we simply take $\ket{\psi_0}=\ket{0}^{\otimes n}$, which requires no additional operations for ansatz preparation. 

It is important to note, both encoder and decoder have hyper-parameters (such as the number of parameters that define the PQCs, the number of neurons and depth of the ANN, etc$.$) for which one must optimise. From here on we will assume that these hyper-parameters are accounted, realising that there is possible future work in rigorously addressing the exact optimisation for this HQA model. 

Now that we have defined the encoder and decoder, the HQA model, $\mathfrak{A}$, is the combination defined as, 
\begin{equation}
    \mathfrak{A}: \mathcal{H}^{\otimes n} \xrightarrow{} \mathcal{H}^{\otimes n} \text{, \ } \mathfrak{A} \big(\ket{\psi_{\text{in}}}\big) =\mathfrak{D} \circ \mathfrak{E} (\ket{\psi_{\text{in}}}) ,
\end{equation}
\noindent where we will refer to the output of the HQA as $\mathfrak{A} (\ket{\psi_{\text{in}}})= \ket{\psi_{\text{out}}}$. Though the HQA looks as though it is a single run through both the encoder and decoder, there is an implicit sub-routine for the encoder where the PQC must be sampled multiple times to obtain $\xi$. 

Now that the components of the HQA have been pieced together, the model is trained to copy the input such that $\ket{\psi_{\text{out}}} \approx \ket{\psi_{\text{in}}}$. 
To do this we find a measure that can identify the distance between quantum states which will be the foundation of the HQA cost function. 
There are many possible ways in which to construct a sensible loss function, the one which we will be considering is one minus the fidelity $\mathscr{F}$ between the model output and the expected training output. Hence for a chosen training data set of $K$ quantum states, $\{\ket{\psi_i^{\text{in}}}\}^{K}_{i=1}$, and the \textit{fidelity} defined as $\mathscr{F}(\ket{\phi}, \ket{\psi}) = \braket{\psi|\phi}\braket{\phi|\psi}$, we define a loss, 
\begin{equation}\label{eq:hqa_loss}
    \mathcal{L}\big(\{\ket{\psi_i^{\text{in}}}\}^{K}_{i=1}\big) = 1 - \widetilde{\mathscr{F}}\Big(\Big\{\mathfrak{A}\big(\ket{\psi_i^{\text{in}}}\big), \ket{\psi_i^{\text{in}}}\Big\}^{K}_{i=1}\Big) , 
\end{equation}
\noindent where
\begin{equation}
    \widetilde{\mathscr{F}}\Big(\big\{ \ket{\phi_i}, \ket{\psi_i} \big\}^{K}_{i=1}\Big) = \frac{1}{K}\sum_{i=1}^K \mathscr{F}\Big(\big\{\ket{\phi_i}, \ket{\psi_i}\big\}^{K}_{i=1}\Big) ,
\end{equation}
 is the average fidelity across all the training instances. In practice, the learner does not calculate the loss over all training instances per iteration, but rather a small \textit{batch} or even over a single instance. This is because, (i) it is computationally expensive having a loss function that sums over all training instances, and (ii) doing so may result in over-fitting to the training data.
 
 The fidelity between two states is maximum when input states are identical, and minimum when they are orthogonal. Hence, the aim is to maximise the fidelity to achieve $\ket{\psi_{\text{in}}} \approx \ket{\psi_{\text{out}}}$, and thereby minimise the loss, which lies in the range $[0,1]$. This is a natural choice for a loss function, as the fidelity is a common distance measure between quantum states. The fidelity has also successfully been used for the construction of denoising quantum autoencoders \cite{10.1103/physrevlett.124.130502} and is hence a great starting point for the construction of the HQA loss function. Selecting a method to measure the fidelity now becomes a hyper-parameter of the model -- for which this paper will use the \textit{swap test} (discussed in Appendix \ref{sec:swap_test}). 

Using the swap test, we can make an estimate of the training complexity. The sampling complexity per iteration (derivation in Appendix \ref{sec:train_compl}) is given by,  
\begin{equation}\label{eq:train_complexity}
    \mathcal{O} \Bigg( \frac{1 + P_{\mathfrak{E}}}{\varepsilon_{\xi}^2}  + \frac{1 + P_{\mathfrak{D}}}{\varepsilon_{\text{fid}}^2}\Bigg)
\end{equation}
\noindent where $P_{\mathfrak{E}}=\text{dim}(\alpha)$ is the number of parameters in the encoder and $P_{\mathfrak{D}}=\text{dim}(\theta)$ in the decoder, $\varepsilon_{\xi}=\Delta \xi_i$ is the uncertainty in each component of the latent vector $\vec{\xi}$, and $\varepsilon_{\text{fid}}$ is the uncertainty in the fidelity measurement. The required $\varepsilon_{\xi}$ and $\varepsilon_{\text{fid}}$ is quite non-trivial. This non-triviality will become evident when dealing with the application in Section \ref{sec:application}, where the HQA is fundamentally unable to learn some states due to their stochasticity.

\subsection{Order in latent space}\label{sec:order_in_lat_space}

Now that the HQA has been constructed, one can observe the powerful nature of representing states in a classical latent space. Training the HQA gives rise to order in latent space that is created purely through matching the input quantum state to the output. In other words, even though we are not supplying the HQA with information about the states trained directly, the model is able to learn these differences and form patterns in latent space.
It is this order that we can exploit to apply ML learning techniques to cluster and classify states in Section \ref{sec:application}.

The HQA is trained for a training set of quantum states $\{\ket{\psi_i^{\text{in}}}\}^{K}_{i=1}$ that have underlying symmetry. In this paper, $\ket{\psi_i^{\text{in}}}$ will correspond to a set of distinct amplitude encoded Gaussian states that can be easily analysed.  

A Gaussian distribution is defined as,
\begin{equation}
    \mathcal{N}(x;\mu, \sigma) = \frac{1}{\sigma\sqrt{2\pi}}e^{-\frac{1}{2}\big(\frac{x-\mu}{\sigma}\big)^2} ,
\end{equation}
\noindent where $\mu$ and $\sigma$ are the mean and standard deviation respectively. Quantising this function for $N$ equally spaced values in the range $\mu \in \big[-\frac{N}{2},\frac{N}{2}\big]$, $\sigma \in \big(0,\frac{N}{3}$\big], we have $|{d}_i(\hat{\mu}, \hat{\sigma})| = \mathcal{N}(i-\big\lceil\frac{N}{2}\big\rceil; \hat{\mu}, \hat{\sigma})$ where $i \in \{0,1,...,N-1\}$ and $\lceil a\rceil$ is the ceiling function that returns the smallest integer above or equal to $a$. Now that we have discrete distribution, ${d}_i(\hat{\mu}, \hat{\sigma})$, it can be encoded into the $N=2^n$ amplitudes of a $n$-qubit quantum state, 
\begin{equation}
    \ket{\psi_{\mathcal{N}} (\hat{\mu}, \hat{\sigma})} = \frac{1}{\sqrt{C}}\sum_{i=0}^{N-1} \hat{d}_i(\hat{\mu}, \hat{\sigma}) \ket{i}, 
\end{equation}
\noindent where $C = \sum_{i=0}^{N-1} |{d}_i(\hat{\mu}, \hat{\sigma})|^2$ is the normalisation constant. The ability for such encoded states to be variationally encoded has been shown in \cite{10.1038/s41534-019-0223-2}.


\begin{figure}
    \centering
    \includegraphics[width=239pt]{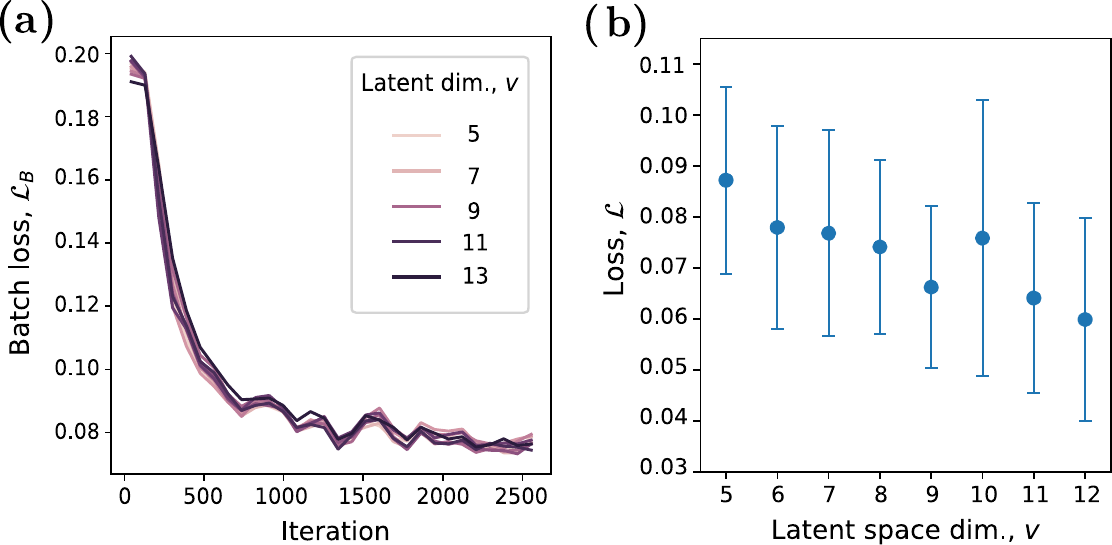}
    \caption{An ensemble of 12 HQA models, with a 5 qubit input state, were averaged for each data point in the two figures. \textbf{(a)} Illustration of the running loss as a HQA was trained with identical samples for various latent sizes. A bin size of 100 was used for plot. \textbf{(b)} Performance of the HQA as the latent space dimensionality is increased. It is important to note that \textbf{(b)} shows the average Loss of 1000 testing instances sampled from a continuous range, $\hat{\mu} \in [\frac{-N}{2}, \frac{N}{2}]$ and $\hat{\sigma} \in [0, \frac{N}{2}]$. Whereas the loss displayed in \textbf{(a)}, $\mathcal{L}_{\text{B}}$, is the running loss during training, averaging over a batch size of 2.  }
    \label{fig:lat_space_anal}
\end{figure}

With this set of distinct quantum states, the HQA is employed to generate a classical latent space that represents the subset in which these states lie. Automatic differentiation is assumed through this unsupervised model to update the PQC parameters in the encoder and the weights of the ANN that determine the decoder. The PQC architecture of $\mathcal{U}_1(\alpha)$ and $\mathcal{U}_2(\theta)$ are shown in Appendix \ref{sec:pqc_arch}.  We take dim$(\alpha) = 4v$ and dim$(\theta) = 4n$, and a feedforward ANN with one hidden layer of size $2v$.
The training was performed in batches of 2 states using the Adam ($\gamma = 0.1$) optimiser \cite{adam_opt}, for a fixed number of epochs -- which is the number of iterations each training sample is used for optimisation. 

In Figure \ref{fig:lat_space_anal}a, we see that training convergence is robust to latent space dimension, with similar loss evolution for all $v$. However, in Figure \ref{fig:lat_space_anal}b, larger $v$ is seen to decrease the loss in testing. As expected we see greater expressibility of the quantum state with a larger latent space. 

\begin{figure}
    \centering
    \includegraphics[width=240pt]{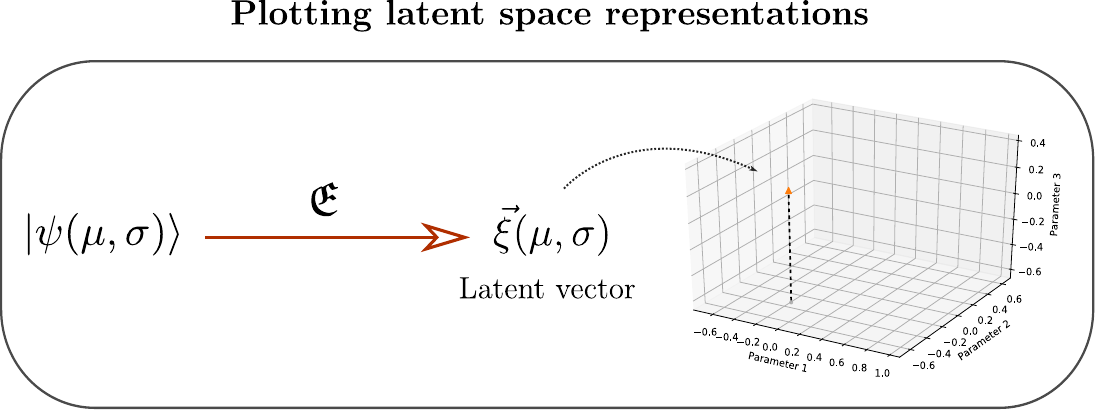}
    \caption{Once trained, one can use the encoder,  to obtain the classical representation of a given quantum state: $\mathfrak{E}(\ket{\psi(\mu, \sigma)}) = \vec{\xi}(\mu, \sigma)$. This latent vector of dimension $v$, can subsequently be plotted using a subset its components. }
    \label{fig:3dplot_latent_rep_process}
\end{figure}

\begin{figure}
    \centering
    \includegraphics[width=234pt]{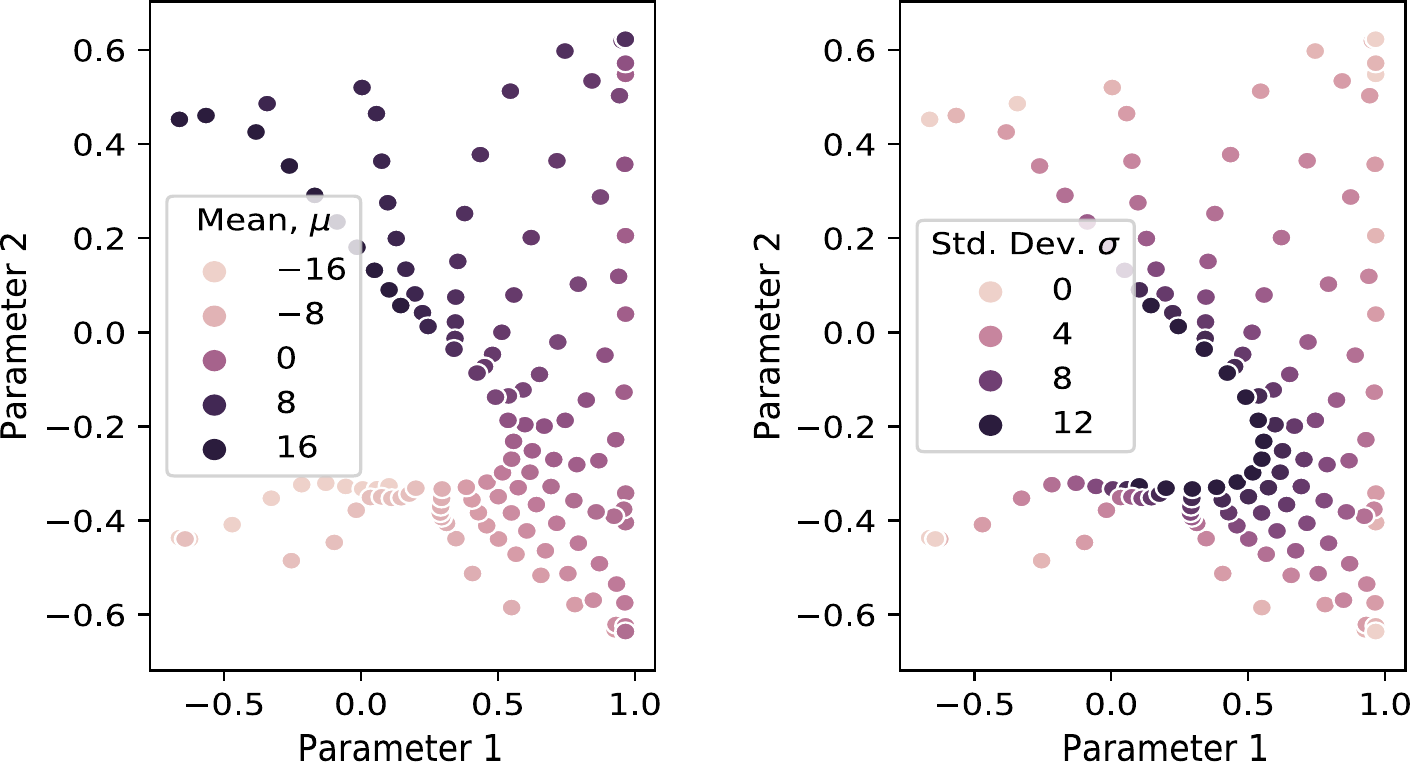}
    \caption{Figures show a HQA trained with Gaussian distributions over 5 qubits and a latent space dimension, $v=12$. Noting that the HQA is not explicitly supplied with information about the states, the model can be seen to decode states into latent space, such that patterns are formed based on aspects of the trained distributions. Here we observe patterns formed in latent space distinguishing the mean and standard deviation of the Gaussian amplitude encoded states.}
    \label{fig:mean_std_latentspace}
\end{figure}

\begin{figure}
    \centering
    \vspace{-0.5cm}
    \includegraphics[width=210pt]{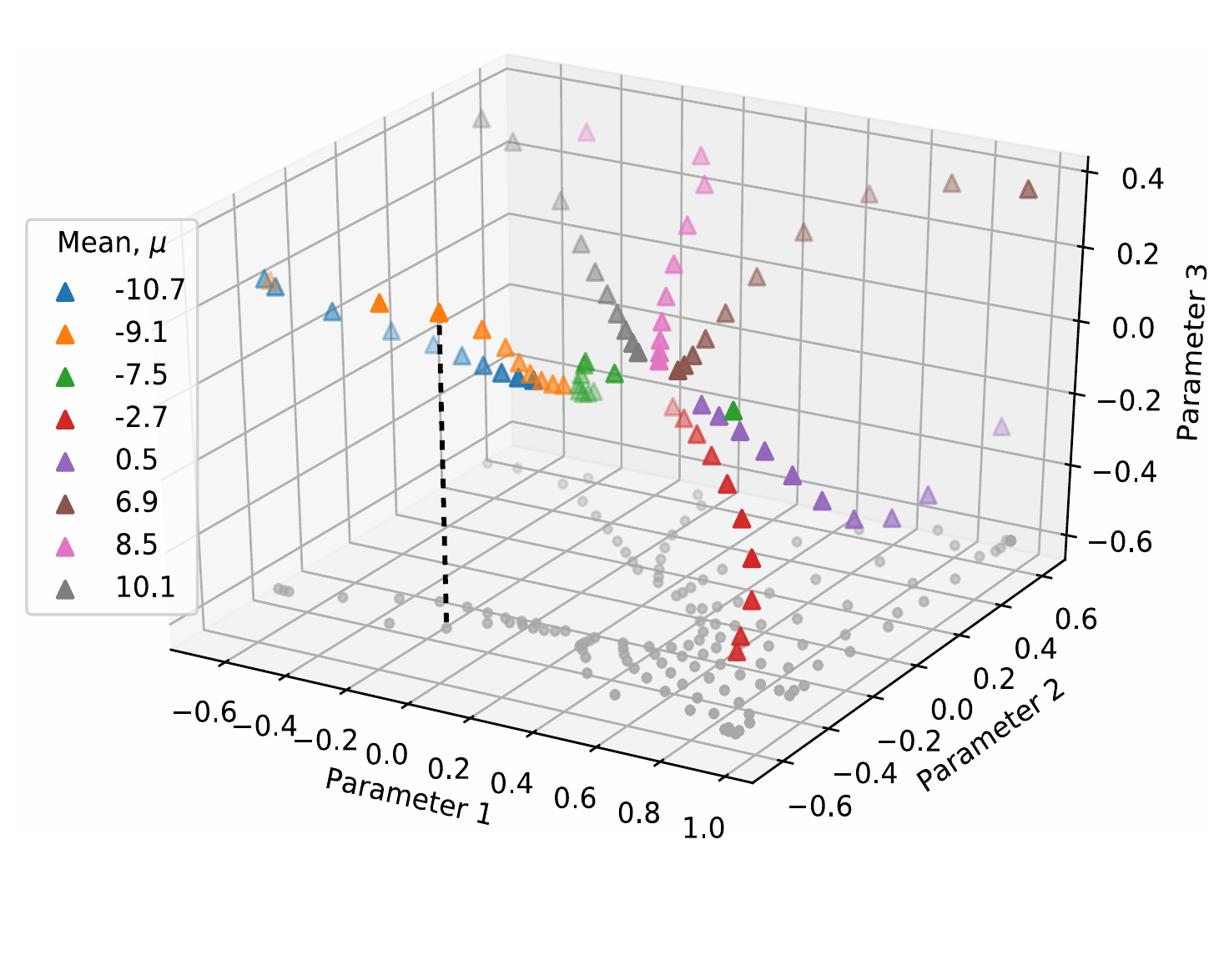}
    \vspace{-0.9cm}
    \caption{ Illustration of the Gaussian distributions displayed with 2 parameters in Figure \ref{fig:mean_std_latentspace}, extended to a $3^{\text{rd}}$ dimension. Its projection onto Parameter 1 \& 2 space is shown in grey. Note: only some of the states with chosen means are plotted.}
    \label{fig:3dplot}
\end{figure}

Once trained, to analyse this model we can apply only the trained encoder, $\mathfrak{E}$, to an input quantum state and observe its location in latent space -- this latent vector is what we refer to as the classical representation of the input quantum state. An illustration of this process is shown in Figure \ref{fig:3dplot_latent_rep_process}. The latent vector can then be obtained for all the amplitude encoded Gaussian states used for training, and then plotted. Figure \ref{fig:mean_std_latentspace} shows such a plot for $n=5$ and a latent size, $v=12$ where first two components (or parameters) of the vector are illustrated. This figure shows that Gaussian encoded states are distributed with a pattern distinguishing their mean and standard deviations. In this specific example, we see that an outward radial movement in the presented latent space corresponds to decreasing the standard deviation; and a positive polar rotation corresponds to increasing the mean. Such elegance is evident from patterns in the original set of quantum states; the two degrees of freedom: $\mu$ and $\sigma$. However, these patterns have the ability to be extremely non-trivial to visualise and this can be seen when we extend the results of Figure \ref{fig:mean_std_latentspace} to a $3^{\text{rd}}$ dimension as shown in Figure \ref{fig:3dplot}. Hence this non-triviality suggests the use of ML to learn patterns in latent space.

It is not necessarily true that patterns will be \textit{visible} when plotting the first two parameters of latent space. A plot of, say, the $9^{\text{th}}$ and $10^{\text{th}}$ latent parameters shows no patterns at all, as points seem to all congregate on a line or point. This indicates that these latent parameters are not being used to distinguish the quantum states, suggesting that a possible dimensionality reduction is possible for the latent space. This is where one can use \textit{principal component analysis} (PCA) \cite{10.5555/3086952} that will both, allow for a clearer understanding of how the latent space is used, and also transform the space so that the \textit{most principal} components can be plotted.

\begin{figure}
    \centering
    \includegraphics[width=250pt]{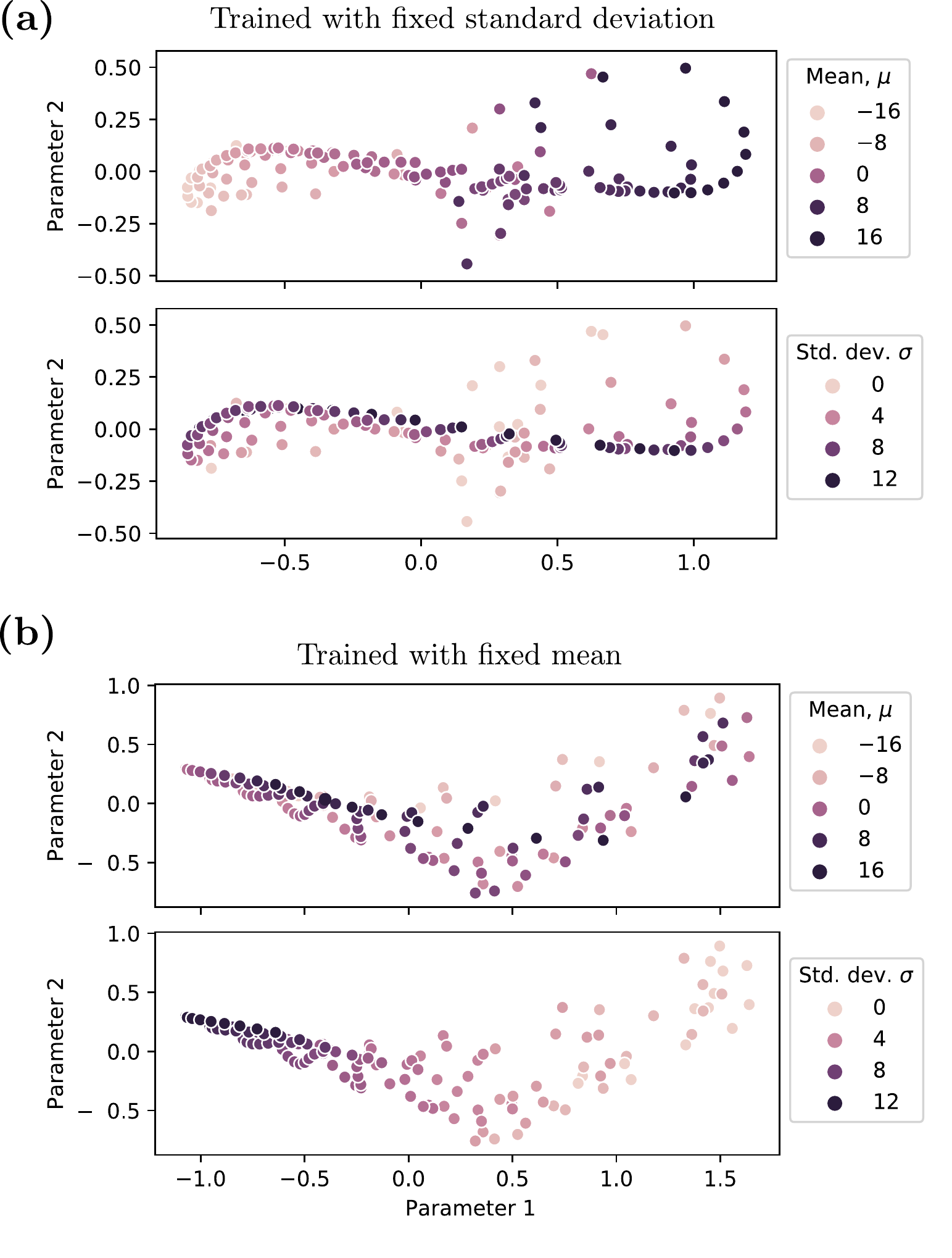}
    \vspace{-0.8cm}
    \caption{\textbf{(a)} The distribution of Gaussian amplitude encoded states in latent space when the HQA is trained with varying mean, but fixed standard deviation of $\sigma=3$. \textbf{(b)} Illustrates the reverse, with training over a range of standard deviations but with a fixed mean, $\mu=0$. Both figures employ a HQA with a latent size of 12 over a 5 qubit amplitude encoded state. The two parameters shown are in the direction of the two most principle components which accounts for $55\%$ of the variation in \textbf{(a)} and $71\%$ of the variation in \textbf{(b)}. This variation refers to the spread of data in the plotted components. One should also note, since the latent space has been transformed to the basis of the 2 principle components, the parameters of the latent space are not necessarily between $-1$ and $1$. }
    \label{fig:train_only_mean_sig}
\end{figure}

\begin{figure*}
    \centering
    \includegraphics[width=500pt]{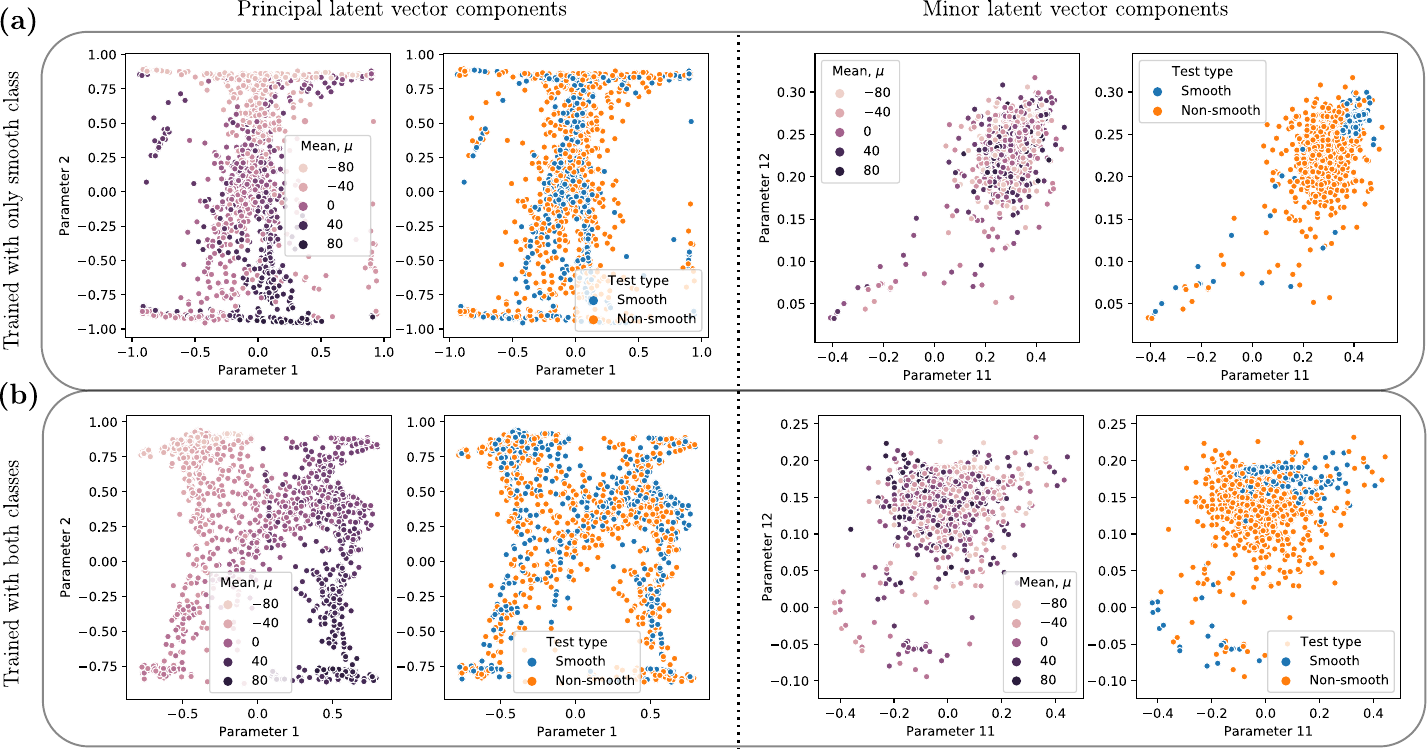}
    \caption{Two HQA models are constructed with 7 qubit input states and a latent size of $v=12$. Figure \textbf{(a)} shows the \textit{actual} location of states in latent space when the HQA model is trained with only \textit{smooth} states of skewed Gaussians. While \textbf{(b)} shows a similar result, however, in this case where the HQA is trained on both classes. Both \textbf{(a)} and \textbf{(b)} include plots for the principal and minor components of the latent vector. In total, 1600 states are plotted, split evenly between the classes.}
    \label{fig:clustering_figures}
\end{figure*}

Now that we have identified a method of systematically analysing latent space, we can ask the question of whether this pattern would still occur if one simply trained on Gaussians with different mean values. Or similarly, if the model was built on training on Gaussians with only varying standard deviations. The results of doing so are shown in Figure \ref{fig:train_only_mean_sig} where the two most principal components from PCA dimensionality reduction are plotted. The formation of patterns in latent space can only be observed when training has \textit{seen} the variations in state. For example, in Figure \ref{fig:train_only_mean_sig} where only $\mu$ was varied and $\sigma$ was kept constant, there is order formed distinguishing $\mu$ but not $\sigma$. The opposite occurs where the training is switched. This suggests the HQA naturally attempts to allocate areas of similar quantum states without the need for additional supervision. In the context of applications, this is extremely useful as one can exploit the latent space location of a particular unknown state in reference to other known states -- where similarity was before not necessarily obvious. In general, this means that we can infer information about states by applying ML algorithms to the states in latent space, as will be explored in Section \ref{sec:application}.

\section{Application}\label{sec:application}

It is evident that clear patterns emerge in latent space from training Gaussian distributions.
An extension to merely observing these patterns is obtaining information about quantum states through their latent representation. This includes both, understanding what it means for states to be located near each other in latent space, and also seeing if one can infer information about states using ML on their latent space representation. 

In order to test these methods, we construct a toy problem involving two classes of states. We define amplitude encoded skewed Gaussian distributions of the form, 
\begin{equation}\label{eq:fuzz_distr}
    {d}_i = {\mathcal{N}}_i \cdot {v}_i \cdot \hat{\eta} , 
\end{equation}
\noindent where $d_i$ is the amplitude of the $i^{th}$ orthogonal state, ${\mathcal{N}}_i$ is a Gaussian distribution, ${v}_i = \max\{0, a\cdot i + b\}$ is a linear function, and we have the class label definitions, 
\begin{equation}\label{eq:class_labels}
\hat{\eta} = 
\begin{cases}
    1 \text{, \ \ class = "smooth"} \\ 
    \eta \text{, \ \ class = "non-smooth"}
\end{cases}
\end{equation}
\noindent where $\eta \in [0,1]$ is a uniform stochastic term that fluctuates as a distribution is called for training or testing. 

To illustrate the power of the HQA, we introduce the artificial objective of clustering states by either \textit{smooth} or \textit{non-smooth}, simply from applying classical ML techniques to their latent space representations. However, before one can obtain these representations, one needs to train the HQA, raising the question of selecting training instances. In general, there is no clear answer to how one should proportion the training instances between the two classes. However, it was seen in Section \ref{sec:order_in_lat_space} that order was formed when the HQA had \textit{seen} different distributions, without which the latent representations appear to not separate deferring distributions. Hence we look at two HQAs: (i) one that is trained with only \textit{smooth} states, and (ii) a HQA trained with both classes in equal proportion. An important point to know here is that the HQA will fundamentally not be able to reproduce the states with the applied stochastic term as the set of such distributions is far too large. Nevertheless, it will be shown -- through both clustering and classification -- that the HQA does not need to recreate states perfectly to be useful in the context of ML in latent space.  

\begin{figure*}
    \centering
    \includegraphics[width=500pt]{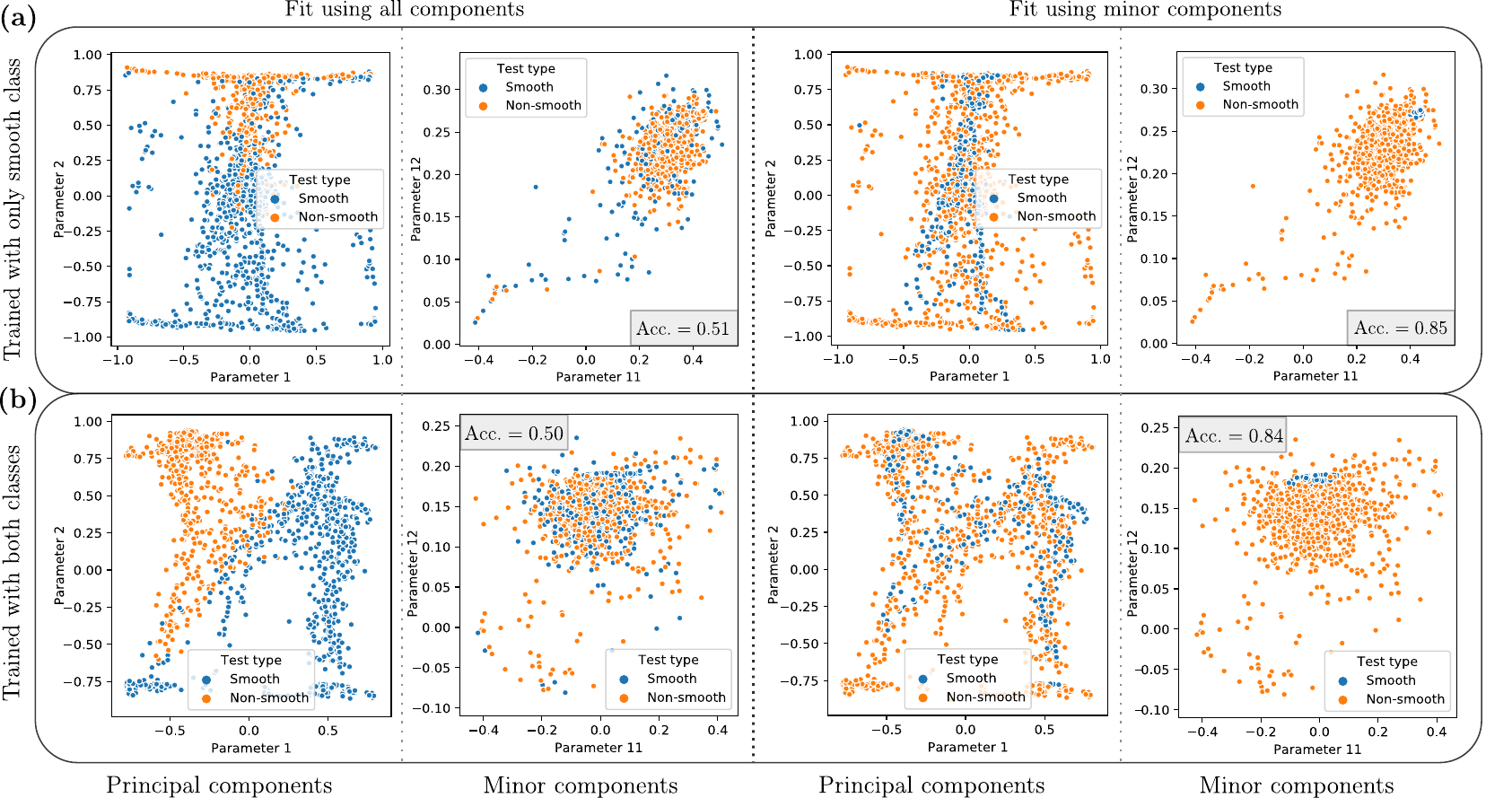}
    \caption{These figures show the prediction results of clustering on the latent space representations of quantum states. The classical clustering was done using Gaussian Mixture Modelling, post the training of the HQA. The clustering fits were taken using: (i) all components and, (ii) the last $4$ minor components (least principal components). The plot shows four possible clustering outcomes by also distinguishing the HQA model trained with only smooth states in \textbf{(a)}, and the HQA trained with both classes in \textbf{(b)}. Each of the four outcomes, have plots showing the predictions in both the principal and minor components. Note: the class labels of \textit{smooth} and \textit{non-smooth} are added later, as clustering merely predicted two groups of blue and orange.}
    \label{fig:pred_labelled_states_lat_space}
\end{figure*}

Post the HQA training, it is possible to analyse the latent representations of the quantum states that are obtained by applying $\mathfrak{E}$ on a sample of the states (the results of which are shown in Figure \ref{fig:clustering_figures}). 
Using both classes of distributions for training, constructs a HQA (Figure \ref{fig:clustering_figures}b) that is still able to allocate regions of varying mean in its principal components. However, the HQA trained on only smooth states lacks this order (Figure \ref{fig:clustering_figures}). Interestingly, minor components of the vector clearly distinguish the two classes of states, regardless of the its training. 

Having constructed classes of states, in this section we will look at (i) clustering states based on their latent representations, and (ii) providing an enhancement on classifying quantum states with semi-supervised learning.

\subsection{Clustering}

In classical ML the most common algorithm for clustering is \textit{kmeans} \cite{10.1109/tit.1982.1056489} (also referred to Lloyd's algorithm) which attempts to find clusters in a data set by observing some classical distance measure (an introduction to kmeans is presented in Appendix \ref{sec:clustering}). The quantum equivalent was first proposed in \cite{qml_lloyd}, in the form of a so-called \textit{quantum kmeans}. Here they are able to produce a quantum state corresponding to the $k$-clusters with complexity that grows linearly with the number of qubits, $n$. However, obtaining a classical description becomes exponential as the quantum states need to be measured using \textit{quantum state tomography} (QST) techniques, such as compressed sensing \cite{10.1103/physrevlett.105.150401, 10.1088/1367-2630/14/9/095022}, which requires $\mathcal{O}(kN^2 \log N)$ where $N=2^n$. 
In addition, the authors exploit the fact that kmeans can be expressed as a quadratic programming problem which can be solved using a quantum adiabatic algorithm. In \cite{10.5555/2871393.2871400}, an approach is presented with efficient quantum methods of calculating the Euclidean distance. Alternatively, the quantum approximate optimisation (QAOA) algorithm \cite{farhi2014quantum} is used in \cite{otterbach2017unsupervised}, for clustering by association to the \textit{maximum cut} problem. 

Many of the methods in literature suffer from requiring QST techniques to classically obtain clusters, which is not required when using the HQA. We define a method of clustering states using classical representations generated by the HQA, after which classical clustering algorithms are used on a space that is exponentially smaller. 

For the classical clustering of latent space, it is required that a more sophisticated method than kmeans is used. There are many advanced clustering methods that allow non-linear clustering \cite{nonlinear_clustering}, but in general such clustering is not natural. In this work we will use \textit{Gaussian Mixture Modelling} which is an extension of the kmeans clustering algorithm that has the flexibility to change the importance of certain parameters of the vector \cite{10.5555/795665.796496} -- kmeans simply uses a Euclidean measure of distance. 

The predicted labels from clustering latent space are shown in Figure \ref{fig:pred_labelled_states_lat_space}. The results show remarkable agreement with true labels when considering the minor components of the latent vector: 85.1\% accuracy for the HQA trained with only smooth states and 84.4\% accuracy for the HQA trained with both classes. On the other hand, clustering based on the principal components are seen to amount to guessing the class of state. The reason for this is evident when looking at the principal components. Fitting using all components is seen to identify clusters relating to the mean of the distributions (grouping negative and positive means), which is also a valid clustering process. Importantly this is only seen for the HQA trained with both classes. Such a deficiency is attributed, not to a limitation of the algorithm, but rather to the non-uniqueness of the task's solution.

At this point, one should retrace the steps of this clustering method in the context of an application. It is conceivable that a quantum experiment is conducted that produces quantum states about which the user has no information. This stream of states could hence be used for the training of the HQA. Importantly, however, one should note that a single sample of a quantum state is not sufficient. There is both, the sub-routine for the encoder, as well as the fidelity computation, that requires multiple copies of the state being produced from the experiment. Once a level of convergence has been reached, a clustering algorithm -- such as kmeans or Gaussian Mixture Modelling -- could be used on the latent representations of these states to identify possible groups. Finally, one can learn to identify -- possibly highly non-trivial -- distinctions between quantum states. This was shown in the distinction between \textit{smooth} and \textit{non-smooth} states, however more work is required to extend such a method to further applications.

\subsection{Semi-supervised classification}

\begin{figure}
    \centering
    \includegraphics[width=240pt]{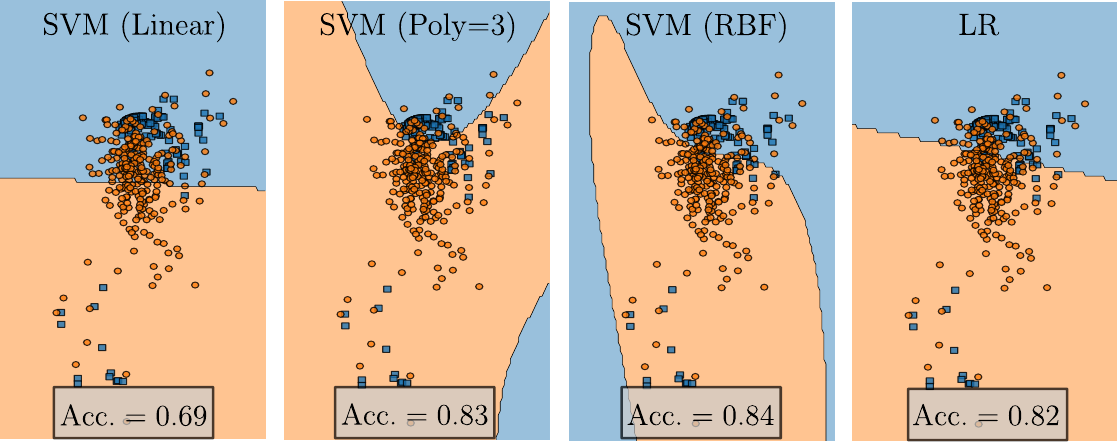}
    \caption{Decision boundaries of classical ML classifiers: support vector machine (SVM) and logistic regression (LR). The SVM has an additional hyper-parameter, known as the \textit{kernel}. Three different kernels are also shown: linear, polynomial (in this case degree of 3) and radial basis function (RBF). For polynomial and RBF kernels, points were fit with $\gamma = 5$. The results shown here are for a subset of the latent points (minor components fitted and plotted) generated from the HQA trained with both classes of states (results from Figure \ref{fig:clustering_figures}).}
    \label{fig:ml_examples}
\end{figure}

\begin{table*}
\centering
\begin{tabular}{c|c|c|c|c}
\hline
ML Model                             & \ Training type \                & Kernel                           & Latent space   & \ Accuracy \  \\ \hline
\hline
\multirow{8}{*}{ \ Support Vector Machine \ }                 & \multirow{4}{*}{Only smooth}     & \multirow{2}{*}{ \ Poly (degree=4) \ }          & \ \ All components \ \  & 0.95      \\ \cline{4-5} 
                                     &                             &                                  & PCA reduction  & 0.55     \\ \cline{3-5} 
                                     &                             & \multirow{2}{*}{RBF} & \ All components \  & 0.92     \\ \cline{4-5} 
                                     &                             &                                  & PCA reduction  & 0.65     \\ \cline{2-5} 
                                     & \multirow{4}{*}{Both} & \multirow{2}{*}{ \ Poly (degree=4) \ }          & \ All components \  & 0.92     \\ \cline{4-5} 
                                     &                             &                                  & PCA reduction  & 0.64     \\ \cline{3-5} 
                                     &                             & \multirow{2}{*}{RBF} & \ All components \  & 0.93     \\ \cline{4-5} 
                                     &                             &                                  & PCA reduction  & 0.75       \\ \hline 
\multirow{4}{*}{Logistic regression} & \multirow{2}{*}{Only smooth}     & \multirow{2}{*}{-}          & \ All components \  & 0.87     \\ \cline{4-5} 
                                     &                             &                                  & PCA reduction  & 0.52     \\ \cline{2-5} 
                                     & \multirow{2}{*}{Both} & \multirow{2}{*}{-}          & \ All components \  & 0.73     \\ \cline{4-5} 
                                     &                             &                                  & PCA reduction  & 0.58    
\end{tabular}
\caption{Results of the binary classification problem of distinguishing \textit{smooth} and \textit{non-smooth} states from their latent space representations. This is done for two HQA models, with $(n,v)=(7,12)$, that identify different latent vectors: a HQA trained on only \textit{smooth} and another HQA trained on both. The accuracy is simply defined as the proportion of correct classifications on a separate group of test data (size=480). Importantly, both the SVM and the LR models are by no means optimised for the performance in this classification problem. Rather, these classical models are used as an illustration of the possibility of classifying states based on their latent space representation. For completeness, the hyper-paramters used for SVM are: $C=1.0$ and $\gamma = 2$ (Linear), $20$ (RBF). Finally, {PCA reduction} takes the 4 most principal components from the latent vector. The PCA reduced space accounts for 55\% of the variance for HQA trained on both classes and 48\% trained only smooth. }
\label{table:ml_in_lat_space}
\end{table*}

In a similar process to using classical clustering methods on latent space, we now use classical supervised learning models to classify quantum states. Specifically two ML algorithms are used: \textit{Support Vector Machine} (SVM) \cite{cortes_support-vector_1995} and \textit{Logistic Regression} (LR) \cite{logistic_regression}. These are both supervised learning algorithms that are successfully used for classification. The former attempts to obtain a separating hyper-plane splitting the smooth and non-smooth classes, while the latter is a form of binary regression. It is enough for the reader to understand that these are supervised learning algorithms that are fundamentally linear, but where the SVM can be extended to non-linear decision boundaries with the use of, what is called, a \textit{kernel}. The decision boundaries of these models are shown in Figure \ref{fig:ml_examples}.

Splitting the $1600$ classical data points (seen in Figure \ref{fig:clustering_figures}) in a $3:10$ testing/training ratio, the SVM and LR models are trained in two ways: (i) on all components of the latent vectors, and (ii) on just the principal components. The accuracy on the test data for all these variations (including kernel) is shown in Table \ref{table:ml_in_lat_space}.

In general, both SVMs with non-linear kernels are seen to have a higher classification accuracy than LR. This makes sense, as the distribution of states on latent space was seen to be highly non-linear. More importantly however, from the performance of the SVM, the interesting result is that the HQA trained on only the smooth distributions, is comparable to the HQA trained on both when considering all components of the vector. At the same time, the HQA trained on both classes performs better when considering ML on only the principal components. To understand this behaviour, we realise that the HQA model attempts to separate states in only a few principal components. This means that when both classes were used for training the HQA, this separation was identifiable by the algorithm. 

Considering all components, the polynomial kernel SVM with a HQA trained on only smooth distributions has a 0.95 accuracy rate. This occurs as a result of the non-smooth distributions -- that were not \textit{seen} by the HQA -- being stored off the learned manifold and into the distinct unutilised regions of the latent space. It was hence easy for both ML algorithms to distinguish between the classes, with even a linear decision boundary from LR achieving an accuracy of 0.87.

These are significant results to keep in mind, however, not necessarily the most natural use of the HQA. If such a classification of quantum states was the main objective given a set of labelled states, one could instead -- potentially -- train the encoder $\mathfrak{E}$ rather than an entire HQA. However, the power of the HQA comes down to its ability to be trained unsupervised -- ie. without any labelling of the states that are being fed. For example, it is possible to train a HQA with the output states of some quantum experiment without \textit{knowing} anything about the states themselves. Post-training one would only be required to label a {few} states and simply perform classical ML to obtain a working quantum state classifier. The reason that this works lies with the ability of the HQA to \textit{learn} a manifold in which relevant states lie. Such a manifold is far smaller than the space of all states, hence requiring far fewer labelled instances to train.
In ML literature, this process is known as \textit{semi-supervised learning}, where only a portion of instances are labelled but where the unlabelled instances are also able to help with the overall classification process.

\section{Conclusion}

QML algorithms are yet to conclusively demonstrate advantage in the NISQ-era. There remain crucial problems that must be resolved for the application of these methods on real devices. One such problem is the emergence of barren plateaus and narrow gorges in gradient-based optimisation. This is common to all models that involve the optimisation of PQCs and is hence critical that it is solved for the specific HQA design implemented in this work. Yet another impediment is the difficulty of obtaining continuous values from qubit-based devices with measurements that are fundamentally digital. In this work, this is overcome with repeated measurement. However, it is possible to apply similar principles to Continuous-Variable (CV) Quantum devices \cite{PhysRevLett.82.1784} for a more efficient measurement of continuous latent space vectors. This approach is left for future work, in align with the QML models proposed in \cite{PhysRevResearch.1.033063}. Finally, the effect of noise on the HQA requires further research, with potential robustness seen from training stochastic states. 

Due to these problems, it was important that the proposal of the HQA was made arbitrary in its specific implementation (as illustrated in Figure \ref{fig:general_hqa_fig}c). Therefore, the crucial aspect of this work is the proposed paradigm of \textit{learning} quantum states through the application of ML techniques on their classical representations -- representations that are generated through training a hybrid quantum autoencoder (HQA). 

To demonstrate its successful application, the HQA was constructed using PQCs on a training set of Gaussian amplitude encoded states. Patterns -- associated with the mean and standard deviation of the Gaussian encoded quantum state -- were visually recognisable in their latent space representations. The emergence of order in latent space was exploited for the implementation of clustering and semi-supervised classification. In the context of \textit{(non-)smooth} states (defined in Eq. \eqref{eq:class_labels}), we were able to achieve 84\% accuracy for clustering and 93\% for classification. Though the accuracy is highly problem dependent, the applied states under question had non-trivial distinctions and hence demonstrates the robustness of this clustering and classification approach. 

Finally, it is assumed that an end-to-end application of the HQA will involve a set of training states obtained from supplemental quantum algorithms -- such as the quantum variational eigensolver \cite{10.1038/ncomms5213}. 
In \cite{10.1088/2058-9565/aa8072} the constructed quantum autoencoder is classically simulated to compress ground states of the hydrogen molecule with various $r$. Furthermore, there have been attempts for entanglement classification \cite{10.1038/srep30188, 10.1088/1367-2630/ab783d}, a process which can potentially employ variational methods. Recently, QAEs were suggested for use in low-rank state fidelity estimation \cite{du2021exploring}, for which the structured classical latent space constructed by the HQA could be exploited. These examples motivate further uses of the HQA from the contrived application studied in this work. Nonetheless, the novel paradigm proposed in this paper, lays the framework for a unique approach to extracting information and obtaining underlying structure from sets of quantum states. 

\begin{acknowledgments}

The authors acknowledge the support provided by the University of Melbourne through the establishment of an IBM Network Q Hub at the University. CDH is supported by a research grant from the Laby Foundation.

The large number of simulations required for this work were made feasible through access to the University of Melbourne's High Performance Computer, \textit{Spartan} \cite{spartan}. The HQA implementation was carried out employing the QML framework provided by the \textit{PennyLane} \cite{bergholm2020pennylane} library.

\end{acknowledgments}

\bibliography{Biblio}

\appendix

\section{Artificial Neural Network}\label{sec:ann}

The idea of Artificial Neural Networks (ANNs) was first proposed in 1957 in an attempt to mimic the way in which the human brain processes visual data \cite{rosen_perceptron}. Since then, the use of ANNs has been ubiquitous in many fields ranging from, image classification to numeric calculations in computationally expensive regions of phase-space \cite{10.1093/mnras/staa713}. It is important to note that ANNs are also regularly used in the context of unsupervised algorithms and in all cases, play the ubiquitous role of a function approximator.

\begin{figure}
    \centering
    \includegraphics[width=205pt]{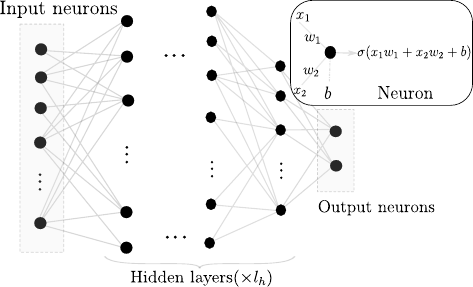}
    \caption{The structure for a feed-forward ANN.}
    \label{fig:ffann_diagram}
\end{figure}

Though ANNs come in various architectures, their fundamental unit is the \textit{neuron} that accept inputs $\vec{x}$ and outputs a scalar known as the activation, $a$, of that neuron. It  has the mathematical form, 
\begin{equation}\label{eq:c_to_q_rep}
a = \sigma\Big(\sum_{i}w_{i}x_{i} + b\Big) = \sigma(\vec{w} \cdot \vec{x} + b),
\end{equation}
where $w_{i}$ is the weighting for input $x_{i}$ and $b$ is a bias term. These parameters will be tuned in the learning stage of the model such that the outputs correspond to the labels of the inputs. The function $\sigma$ is known as the activation function which is a non-linear mapping $\sigma: \mathbb{R} \rightarrow [0,1]$, that outputs the \textit{activation} of a neuron. In practice common activation functions include sigmoid, hyperbolic tan, rectified linear unit (ReLU) functions \cite{10.5555/3086952}. 
A \textit{network} of these neurons form an Artificial Neural Network, with the output neurons commonly giving a probability distribution over the possible classifications. There are many different types of architectures for the way in which the neurons can be connected, but a common example is the feed-forward ANN, shown in Figure \ref{fig:ffann_diagram}. The layered fashion of the feed-forward ANN means that the whole network has a relatively simple mathematical form,
\begin{equation}\label{eq:ffann_math_rep}
    f(\vec{x}) = \sigma_{L} \circ g_{L} \circ \sigma_{L-1} \circ g_{L-1} \cdots \sigma_{1} \circ g_{1}(x),
\end{equation}
\noindent where we have, $g_{i}(\vec{x}) = \boldsymbol{W}_{i}\cdot \vec{x} + \vec{b}_{i}$ is the sum of previous layer neurons appropriately weighted by matrix $ \boldsymbol{W}_{i} \in \mathbb{R}^{l_{\text{in}}} \times \mathbb{R}^{l_{\text{out}}} $. These weights are, most commonly, learned through gradient optimisation. 

The real utility of ANNs is in their ability to learn and approximate any continuous function with sufficient data. 
The \textit{Universal Approximation Theorem} states that a feed-forward neural network with a single hidden layer is able to approximate any continuous function on $\mathbb{R}^{N}$ \cite{10.1109/72.363453}. It is important to note that non-linear activation functions are required -- without which we have a complex linear model. This is evident from equation \eqref{eq:ffann_math_rep} where it can be that linear $\sigma_{i}$ implies a linear $f$ as function compositions preserve linearity. The non-trivial aspect is that there is no restriction on the non-linearity of the activation function and how this relates to the number of neurons required in the hidden layer. Due to this universality of function approximation, ANNs do not have the same \textit{hyper-parameter} tuning problem as other ML models.  Hyper-parameters refer to parameters of the model that must be identified by the user -- such as the number of neurons in the case of an ANN. The hyper-parameter that an ANN does not require is a choice of the type of decision boundary, which in other models is crucial. At the same time a significant problem with ANNs its tendency to approximate too closely the training instances rather than generalising upon these data points. This is known in ML literature as \textit{over-fitting} and there are \textit{regularisation techniques} to mitigate this problem. 

\section{The parameter shift rule}\label{sec:param_shift}

\begin{figure*}
    \centering
    \includegraphics[width=480pt]{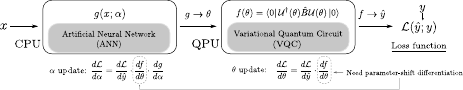}
    \caption{Diagram of an example hybrid QML algorithm, with the combination of an ANN and a PQC. For simplicity, both models only have a single parameter that is trained. For input, $x$, and label, $y$, the diagram highlights the derivatives required for the update of each parameter using the back-propagation algorithm.}
    \label{fig:ml_qml_hybrid_fig}
\end{figure*}

In this paper the computation of quantum gradients of continuous parameters are computed using  \textit{parameter-shift differentiation} proposed in \cite{10.1103/physreva.99.032331}. It is shown that the analytical gradient of a variational circuit, $f$, defined in equation (\ref{eq:variational_circuit}), can be found when it is composed of gates of the form $\mathcal{G}(\mu) = e^{-i \mu G}$, where they are generated by a Hermitian operator $G$ with strictly two eigenvalues $\pm r$. It is shown that $\partial_{\mu}f$ can be estimated using two additional evaluations of the quantum device by placing either gates $\mathcal{G}(\pm \frac{\pi}{4r})$ in the original circuit next to the gate that we are differentiating. Since for unitarily generated one-parameter gates we have $\mathcal{G}(a)\mathcal{G}(b) = \mathcal{G}(a+b)$, we simply have a shift of the parameters by $s=\frac{\pi}{4r}$ to find the gradient,
\begin{equation}
    \partial_{\mu} f = r [f(\mu + s) - f(\mu - s)].
\end{equation}
\noindent This is aptly named the \textit{parameter shift rule}. For generator $G$ with more than two eigenvalues this strategy fails; however one is able to use an ancilla qubit and perform a decomposition of the derivative of the gate to obtain a gradient as elaborated in \cite{10.1103/physreva.99.032331}. As an important additional benefit, the parameter-shift rule has been shown to hold on noisy quantum devices \cite{Meyer2020}. Furthermore, recent works have attempted to generalise this method of obtaining an analytic gradient for more intricate unitaries using stochastic techniques \cite{10.22331/q-2021-01-25-386}.

Having acquired the derivative of parameters in VQCs, one is now able to explore its use in QML -- especially \textit{hybrid} models involving both classical and quantum processing units. The ability to compute gradients of VQCs means that it is possible to attach a VQC and ANN as components of a larger ML algorithm, over which the back-propagation algorithm will still be applicable (shown in Figure \ref{fig:ml_qml_hybrid_fig}). To implement these models, an open-source \texttt{Python3} software framework for hybrid quantum-classical optimisation and ML, termed \textit{PennyLane} \cite{bergholm2020pennylane}, is used. The library interfaces with popular machine learning libraries such as Tensorflow, PyTorch, autograd while also providing APIs for the access of publicly available quantum devices such as those by {\textbf{Rigetti}} and {\textit{IBM}} -- alleviating some of the tedious programming.

\section{PQC architecture}\label{sec:pqc_arch}

\begin{figure}
    \centering
    \includegraphics[width=200pt]{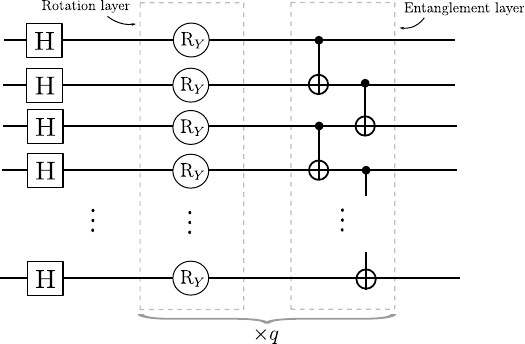}
    \caption{The architecture is defined by alternating rotation and entangling layers -- with $q$ number of such repetitions. Here, $R_{\text{Y}}:= \exp (-iY \theta)$, where $Y$ is the Pauli-Y operator. With only nearest neighbour couplings, such a structure is also known as a \textit{hardware efficient ansatz.}}
    \label{fig:ry_architecture}
\end{figure}

The decomposition unitary $\mathcal{U}(\theta)$ need not be understood in terms of the commonly used single and two qubit circuit gates. Rather, they can be generalised to any tunable parameter in a particular quantum device and optimised accordingly. Though this is possible for specific devices, this paper will address PQCs from a general perspective, such that $\mathcal{U}(\theta)$ is composed of only single and two qubits gates, with parameters $\theta$ defining the rotation on single qubit gates. Though $\mathcal{U}(\theta)$ can, in theory, be set as a random set of U3 gates and CNOT gates, it is generally desired to have a \textit{pattern} of gates so that one can compare and generalise PQCs made from certain patterns. These patterns will be referred to as \textit{PQC architectures}, separate from \textit{device architectures} that describes the physical qubit layout of a device. In this work, we employ an alternating structure of rotation layers and entangling layers, as shown in Figure \ref{fig:ry_architecture}. 

\section{Swap test}\label{sec:swap_test}

\begin{figure}
    \centering
    \includegraphics[width=186pt]{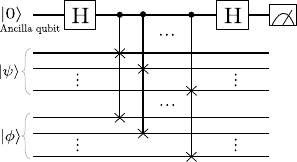}
    \caption{This is a circuit diagram depicting the \textit{swap test} algorithm that is used to compute the overlap, $\lvert \braket{\phi | \psi}\rvert^2$ through repeated measurements of the ancilla qubit.}
    \label{fig:swap_test}
\end{figure}

\noindent To measure the fidelity between two coherent quantum states, one can introduce an ancilla qubit and use the \textit{swap test}. To find the overlap between an output state, $\ket{\psi} = \sum_i \psi_i\ket{i}= \sum_i \ket{\Tilde{\psi}_i}$, and a reference state, $\ket{\phi} = \sum_j \phi_j\ket{j}= \sum_j \ket{\Tilde{\phi}_j}$, one repeatedly measures the ancilla qubit from the circuit shown in Figure \ref{fig:swap_test}. This works since the state before measurement is, 
\begin{align}
    \frac{1}{2} \sum_{i,j} \Big\{ \ket{0}\Big(  \ket{\Tilde{\psi}_i}&\ket{\Tilde{\phi}_j} +  \ket{\Tilde{\phi}_j}\ket{\Tilde{\psi}_i} \Big)  \nonumber \\
    &+ \ket{1} \Big(  \ket{\Tilde{\psi}_i}\ket{\Tilde{\phi}_j} - \ket{\Tilde{\phi}_j}\ket{\Tilde{\psi}_i} \Big) \Big\} .
\end{align}
\noindent Now measuring the ancilla qubit in the Z-basis, the probability of measuring eigenvalue, $z=\pm 1$ is, 
\begin{align}
    \text{Pr}(z=\pm 1) &= \frac{1}{4} \sum_{ijkl} 2 \big( \braket{\Tilde{\psi}_k|\Tilde{\psi}_i}\braket{\Tilde{\phi}_l|\Tilde{\phi}_j} \pm \braket{\Tilde{\phi}_l|\Tilde{\psi}_i}\braket{\Tilde{\psi}_k|\Tilde{\phi}_j} \big) \nonumber \\
            &= \frac{1 \pm \sum_{i,j} |\braket{\Tilde{\phi}_j | \Tilde{\psi}_i}|^2}{2} ,
\end{align}
\noindent where we have used $\braket{\Tilde{\psi}_i|\Tilde{\psi}_j}=|\psi_i|^2\delta_{ij}$ due to orthogonality and the normalisation condition, $\sum_i |\psi_i|^2=1$. Similarly for the $\ket{\Tilde{\phi}_j}$ states. Therefore by measuring and recording the ancilla qubit enough times, we can work out the overlap between the two states in terms of the probability $\text{Pr}(z=\pm 1)$, 
\begin{equation}\label{eq:swap_test_out}
    \mathscr{F}(\ket{\phi}, \ket{\psi}) = \lvert\braket{\phi | \psi}\rvert^2= \mp \Big[ 1 - 2\text{Pr}(z=\pm 1) \Big] .
\end{equation}
\noindent Under certain conditions, such a measure is equivalent to the \textit{mean-squared error} (MSE) between amplitudes encodings, though in general it is a far stronger statement. The MSE Loss, $\mathcal{L}_{\text{MSE}}$, can in this context be written as, 
\begin{equation}\label{eq:mse_equiv_fid}
    \mathcal{L}_{\text{MSE}} = \sum_i \big( |\psi_i| - |\phi_i| \big)^2 \xrightarrow[\text{minimising}]{\text{Equivalent to}} - \sum_i |\psi_i| |\phi_i| ,
\end{equation}
\noindent where again we have made a simplification using the fact that $\psi_i$ and $\phi_i$ are normalised. This means that minimising the fidelity implies the minimisation of $\mathcal{L}_{\text{MSE}}$ -- with the assumption that the amplitudes of the two states are non-negative real values. This can be shown by using the triangle inequality on equation \eqref{eq:swap_test_out} to show that $\lvert\braket{\phi | \psi}\rvert^2 \leq \sum_i (\psi_i \phi_i)^2$ revealing that the maximisation of the fidelity \textit{implies} the minimisation of $\mathcal{L}_{\text{MSE}}$ in equation \eqref{eq:mse_equiv_fid}. However, the fidelity statement is stronger due to the inequality. It should, however, be noted that the swap test requires large numbers of SWAP gates, and hence further research is required to enhance this fidelity measurement. 

\section{Training complexity}\label{sec:train_compl}

Given the particular design of the HQA (proposed in Section \ref{sec:hqa_design}), one can obtain a complexity for the training process.
Each evaluation of the HQA would require $M_{\mathfrak{E}}$ samples to determine the latent vector and $M_{\text{swap}}$ samples to obtain the fidelity. We can relate the error in fidelity, $\varepsilon_{\text{fid}}$, to $M_{\text{swap}}$ such that we have $M_{\text{swap}} = 1/\varepsilon_{\text{fid}}^2$.  This relation relies on the fact that the swap test measurement is a Bernoulli trial. Similarly, we relate the uncertainty in the latent vector components to the number of encoder samples: $M_{\mathfrak{E}} = 1/\epsilon_{\xi}^2$, where $\epsilon_{\xi}$ is the uncertainty in each component of the latent vector. 

A single evaluation of the HQA requires $M_{\mathfrak{E}} + M_{\text{swap}}$ samples. However, training requires the calculation of gradients through the parameter shift rule (Section \ref{sec:param_shift}). Specifically, one must obtain the gradients with respect to all parameters in the model: $\{\frac{d\mathcal{L}}{d\alpha_i}\}$, $\{ \frac{d\mathcal{L}}{dw_{ij}} \}$ and $\{ \frac{d\mathcal{L}}{d\theta_i} \}$, where $\alpha_i$ are the $P_{\mathfrak{E}}$ encoder parameters, $w_{ij}$ are the ANN parameters (weights) and $\theta_i$ are the $P_{\mathfrak{D}}$ decoder parameters. The gradient evaluations are of the form, 
\begin{align}
    \frac{d\mathcal{L}}{d\theta_i} &= \frac{d\mathcal{L}}{d\mathscr{F}} 
    \frac{d\mathscr{F}}{d\theta_i} \\
    \frac{d\mathcal{L}}{d w_{jk}} &= \mathlarger{\sum}_i
    \frac{d\mathcal{L}}{d\theta_i} 
    \frac{df_i}{d w_{jk}}    \\
    \frac{d\mathcal{L}}{d\alpha_l} &= \mathlarger{\sum}_i
    \frac{d\mathcal{L}}{d\theta_i} 
    \frac{df_i}{d\mathfrak{E}}
    \frac{d\mathfrak{E}}{d\alpha_i}
\end{align}
\noindent where the output of the encoder, $\mathfrak{E}$, becomes the input to the ANN, $f$, and the output of the ANN becomes $\theta_i$. Note, only $\alpha_i$ and $w_{ij}$ are optimised, but we also require the gradient with respect to $\theta_i$. 

Sampling a quantum circuit is required to obtain the gradients $\frac{d\mathscr{F}}{d\theta_i}$ and $\frac{d\mathfrak{E}}{d\alpha_i}$. $\frac{d\mathscr{F}}{d\theta_i}$ must be calculated for all $P_{\mathfrak{D}}$ parameters and $\frac{d\mathfrak{E}}{d\alpha_i}$ must be calculated for all $P_{\mathfrak{E}}$ parameters. The number of circuit samples required for $\frac{d\mathfrak{E}}{d\alpha_i}$ and $\frac{d\mathscr{F}}{d\theta_i}$, is proportional to $M_{\mathfrak{E}}$ and $M_{\text{swap}}$, respectively. 
In addition, an evaluation of the HQA is required for the calculation of the loss for that particular iteration. This therefore requires an additional $M_{\mathfrak{E}} + M_{\text{swap}}$ samples. Putting all this together we have sampling complexity per iteration of,
\begin{align}
    \mathcal{O} \Big( (1 + P_{\mathfrak{E}}) M_{\mathfrak{E}} &+ (1 + P_{\mathfrak{D}}) M_{\text{swap}} \Big)  \nonumber \\
    = &\mathcal{O} \Bigg( \frac{1 + P_{\mathfrak{E}}}{\varepsilon_{\xi}^2}  + \frac{1 + P_{\mathfrak{D}}}{\varepsilon_{\text{fid}}^2}\Bigg)
\end{align}

\section{Kmeans clustering}\label{sec:clustering}

Clustering is an unsupervised ML algorithm that attempts to group a set of data points into distinct clusters. The most commonly used clustering algorithm is the kmeans algorithm (or Lloyd's algorithm) \cite{10.1109/tit.1982.1056489} -- used mainly due to its elegance and simplicity. 

Kmeans is an iterative algorithm that partitions a data set into $k$ groups. The algorithm works as follows:
\begin{enumerate}
  \item Select $k$ points randomly from the set to act as seed clusters, $\{\vec{\mu}_i\}_{i=1,...k}$
  \item Assign each data point in the set, $\{\vec{x}_j\}$, to the closest cluster based on some distance measure $D(\vec{x}_j, \vec{\mu}_i)$.
  \item Average the group of points associated with each cluster, to form $k$ new clusters.
  \item Go back to 2, or stop when no reassignments of data points from their cluster groups are made. 
\end{enumerate}

Note that the distance measure, $D$, is arbitrary but usually chosen to be the Euclidean distance. The average of the $m^{\text{th}}$ cluster group, with $M$ associated data points $\{\vec{x}^{(m)}_i\}_{i=1,...,M}$, is taken as $\frac{1}{M}\sum_{i=1}^M \vec{x}^{(m)}_i$.

\end{document}